\pdfoutput=1

\documentclass[3p,times,twocolumn]{elsarticle}
\usepackage{epsfig}
\usepackage{graphicx}

\usepackage{amssymb}

\usepackage[figuresright]{rotating}

\usepackage{amssymb}
\usepackage{subfigure}
\begin{document}
%
\begin{frontmatter}
%
%
%
%
%
%
\title{Test and characterization of a prototype silicon-tungsten \\ electromagnetic calorimeter}
%
%

\author[label1]{Sanjib Muhuri}
\author[label2]{Sourav Mukhopadhyay} 
\author[label2]{Vinay B. Chandratre}
\author[label2]{Menka Sukhwani}
\author[label3]{Satyajit Jena}
\author[label1]{Shuaib Ahmad Khan}
\author[label1]{Tapan K. Nayak}
\author[label1]{Jogender Saini}  
\author[label1]{Rama Narayana Singaraju}

\address[label1]{Variable Energy Cyclotron Centre, Kolkata - 700064, India}
\address[label2]{Bhabha Atomic Research Centre, Electronics Division,
  Trombay, Mumbai - 400085, India}
\address[label3]{Indian Institute of Technology, Bombay, Mumbai - 400076, India}

\begin{abstract}

New generation high-energy physics experiments demand 
high precision tracking and accurate measurements of a large
number of particles produced in the collisions of elementary particles and heavy-ions. 
Silicon-tungsten (Si-W) calorimeters  provide the
most viable technological option to meet the requirements of particle
detection in high multiplicity environments. 
We report a novel Si-W calorimeter design, which is optimized for
$\gamma/\pi^0$ discrimination up to high momenta.
In order to test the feasibility of the calorimeter, a prototype mini-tower
was constructed
using silicon pad detector arrays and tungsten layers.
The performance of the mini-tower was tested using pion and electron beams at the
CERN Proton Synchrotron~(PS). The experimental results are compared
with the results from a detailed GEANT-4 simulation.
A linear relationship between the observed energy 
deposition and simulated response of the mini-tower has been obtained,
in line with our expectations.

\end{abstract}

\end{frontmatter}

\section{Introduction} 

One of the major challenges in high-energy physics experiments 
is to detect and analyze majority of the particles produced
in the collisions of elementary particles or heavy-ions.
Experiments at the CERN Large Hadron Collider (LHC) and at the planned
colliders (International Linear Collider or Future Circular
Collider) necessitate accurate energy measurements over a large dynamic range and high precision 
tracking of the particles emitted in the collisions.
Compared to proton-proton collisions, the challenges get manifolded for the heavy-ion 
(such as Pb-Pb) collisions due to the increase in the number of emitted particles
by several orders of magnitude.
A calorimeter~\cite{Calorimeter-1} is normally used for accurate characterization 
of an incoming particle in terms of its position and energy. 
Sampling calorimeters, segmented in longitudinal as well as in transverse
directions, are considered to be ideal for accurate
characterization of particles in terms of their positions and energies.

In this article, we consider Si-W calorimetry, using
tungsten as converters/absorbers and segmented silicon pad
detectors as active medium.
Because of good energy and position resolution characteristics,
silicon detectors are considered to provide the most suitable detection
medium. Although
the silicon pad detectors used here have no intrinsic charge gain, they exhibit
good charge collection efficiency, fast response time and require low
bias voltage to achieve full  depletion.
Low sensitivity of silicon detectors to the
magnetic field enables the calorimeter to be compatible with 
complex experimental setup in high magnetic field environment. 
At the same time, use of tungsten as absorber makes the electromagnetic 
calorimeter relatively compact and helps to achieve full energy containment.
Combination of silicon and tungsten layers offer one of the most novel design 
for calorimetry in high energy
physics experiments~\cite{rancoita-1,rancoita-2,sicapo-1,sicapo-2,sicapo-3,
  wizard, opal, calice-1,calice-2,phenix,phenix1,strom,cluster,Calorimeter-2}.

In this article we report the design of a Si-W calorimeter,
which can provide accurate position 
and energy measurement with minimum leakage in both  transverse and longitudinal directions.
In order to test the feasibility of such a design, 
a mini-tower, consisting of four layers of silicon pad
detectors and tungsten plates as absorbers, has been constructed.
The design, performance and characterization of the
mini-tower has been discussed in detail.
Though the prototype is limited with a few course layers, the
basic characteristics, like, longitudinal shower profile, calibration
of energy deposition, etc., can aid to build the full length calorimeter.
The article is organized as follows. Section~2 gives the
design of the Si-W calorimeter and in section~3,
setup of the mini-tower is discussed. Details of the
silicon pad detector arrays with readout electronics are discussed in
section~4. In section~5, 
the laboratory test results of silicon pad detector with
$\beta$-source is presented.
The test beam setup is given in section~6.
The results of the energy deposition from simulated and experimental
data 
have been 
described and compared in section~7. The article is concluded with a
summary in section~8.

\section{Design of a Si-W Calorimeter}

A sampling type electromagnetic calorimeter with silicon detector 
as active medium and high purity tungsten as absorber
has been simulated using GEANT4 toolkit~\cite{geant4_1,geant4_2}.
The main goal is to optimize the design with adequate energy and position
resolutions in order  to handle extreme high particle density
environment. The design consisted of
20 layers of silicon pad arrays and tungsten combinations. 
Each layer consisted of a 3.5~mm thick tungsten plate 
followed by a 0.3~mm thick silicon sensor. 
In the simulation, pure tungsten with density of 19.3 gm/cm$^3$ has
been used. 
Being a high-Z element, tungsten converts high energy photons or
electrons into electromagnetic showers. The majority of the photons that
are emitted in high energy collisions are decay photons from
$\pi^0$ mesons. One of the major goals is to reconstruct $\pi^0$
mesons and their
energy from the measured photon showers. The decay angle of the two emitted
photons decreases with the increase of the energy of $\pi^0$ mesons. 
The reconstruction of
the photon showers needs to be accurate in order to obtain the shower
positions of the photons and deposited energy. In order to measure
$\pi^0$ 
energy accurately, tracking of the shower in different layers
is needed. Therefore, the good position resolution of the detectors in the
sensitive medium is essential.

\begin{figure}[tbp]
\includegraphics[scale=.4]{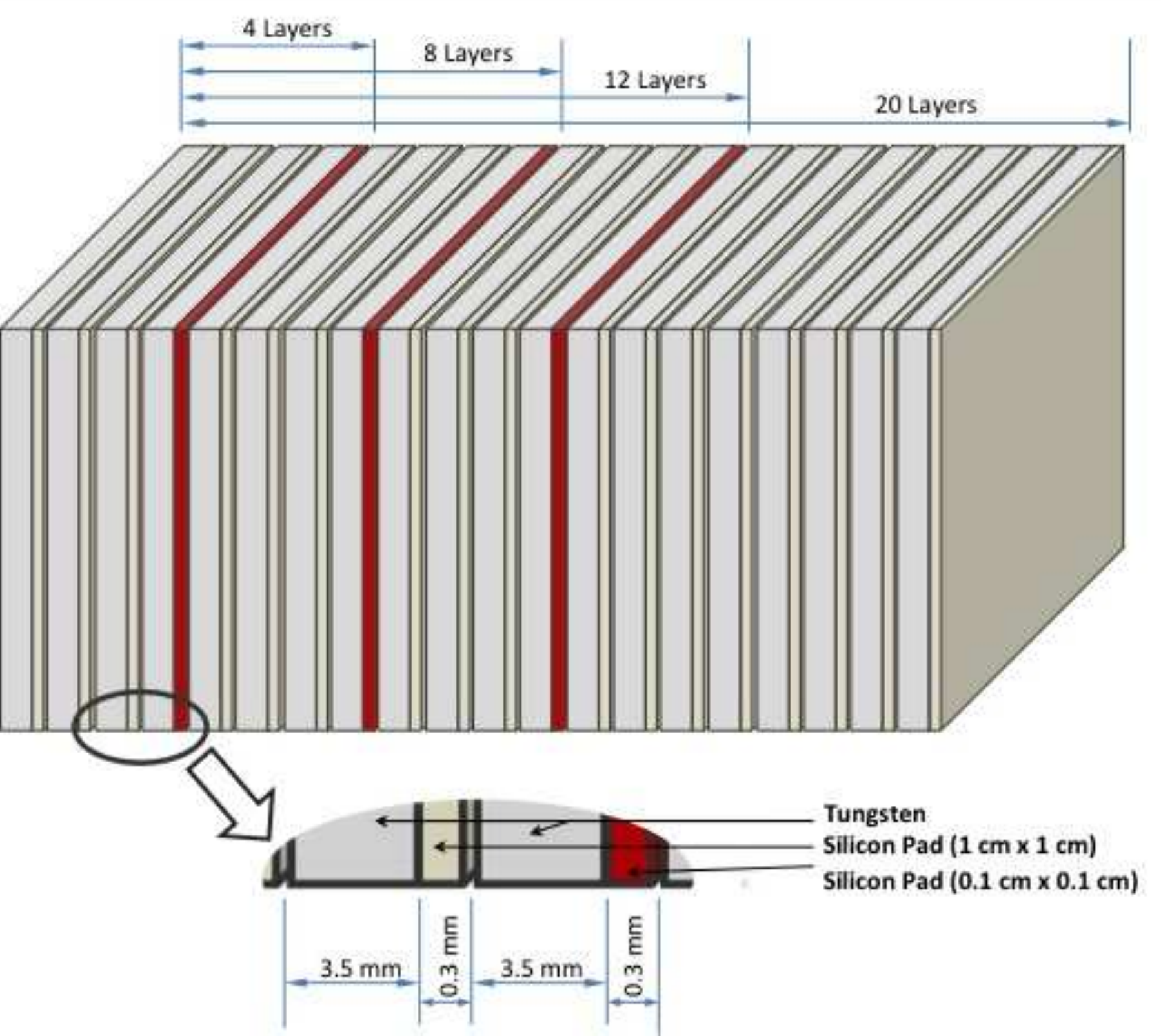}
\caption{The components of a tungsten-silicon sampling calorimeter consisting of 
20 layers of detectors. The three highlighted layers 
are made of highly granular 0.1~cm~$\times$~0.1~cm silicon pads, 
and the rest of the layers consist of 1 cm x 1 cm silicon pads.}
\label{fig:calorimeter}
\end{figure}

The calorimeter,  shown in Figure~\ref{fig:calorimeter}, was designed 
considering all of the above requirements into account~\cite{cluster}. 
 Three high granular planes were placed in the region of the shower maxima for accurate measurements.
Twenty sensitive detector layers were considered, 
out of which three layers (4,
8 and 12) were made of highly granular silicon pads, each with dimension of
0.1 cm $\times$ 0.1 cm.
The rest of the layers were made up of
1 cm $\times$ 1 cm silicon pad detectors. The three high granular planes
help to determine the shower position with high accuracy and thus help
in tracking the path of the incoming particles. 
Reconstruction of clusters (produced from electromagnetic showers of 
electrons or photons at consecutive layers) using the Fuzzy Clustering
technique was carried out and found to
be quite powerful in differentiating overlapping clusters~\cite{cluster}.
Responses of single particle events (electrons, photons, $\pi^{\pm}$
and $\pi^0$) were obtained by placing the
calorimeter at a distance of 400~cm from the interaction point. 
The performance of the calorimeter has been found to be efficient in
discriminating $\gamma$ to $\pi^0$ up to ~200~GeV.
The feasibility of the calorimeter has further been studied by
constructing a mini-tower and testing it with
pion and electron beams of various energies.

\section{Mini-tower arrangement}

\begin{figure}[tbp]
\centering
\includegraphics[width=0.97\linewidth]{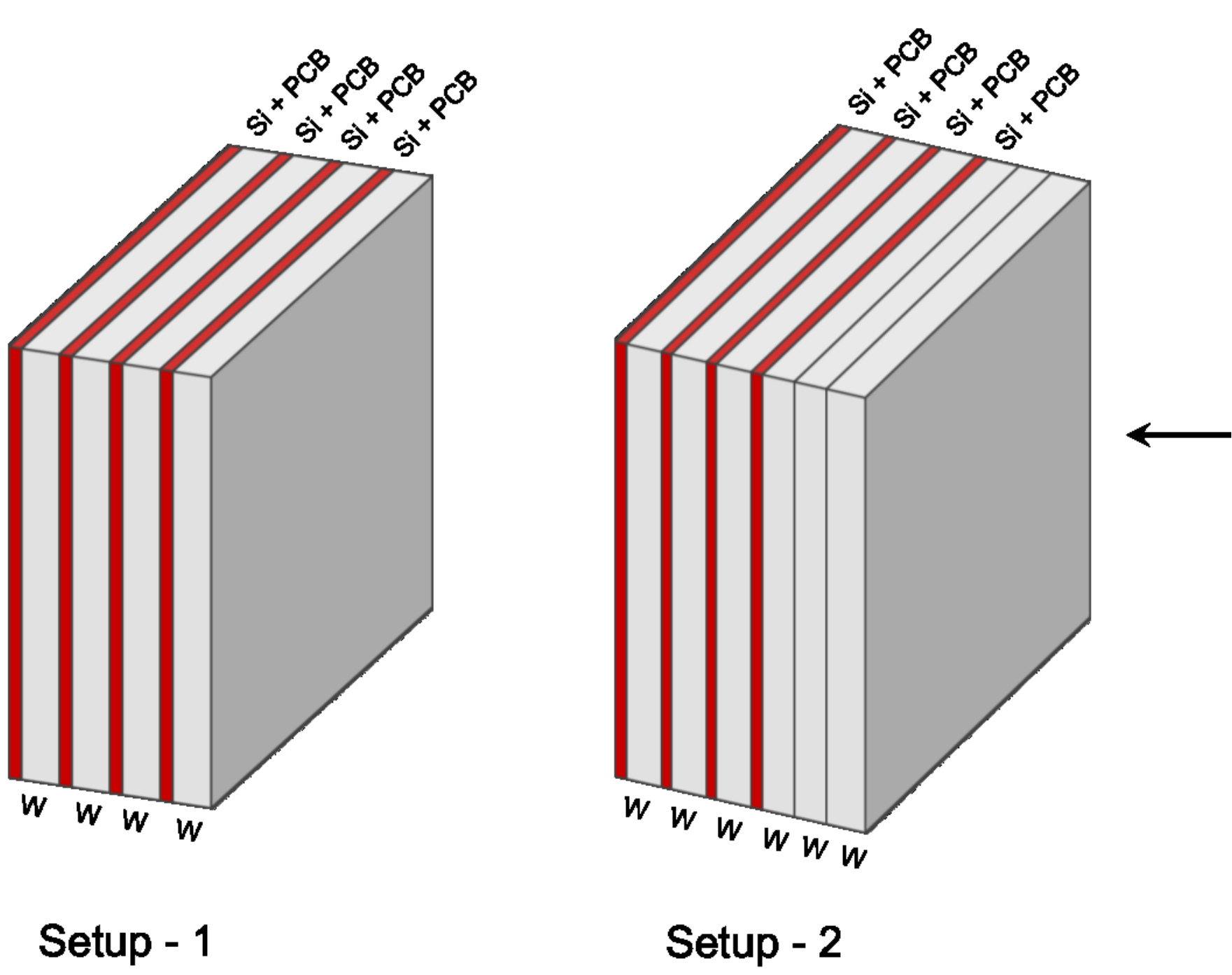}
\caption{ Two different setups of the mini-tower arrangement used
  for the test at CERN-PS.}
\label{minitower}
\end{figure}

The prototype tests of the Si-W calorimeter were performed
with the help of
two  different setups of the mini-tower by using
four layers of silicon pad detector arrays with readout electronics
and several tungsten plates.  Each of the tungsten plates is 1~radiation length ($X_0$) thick and with
99.9\% purity. 
The setups are considered by
keeping in mind the more realistic case to be implemented in future experiments.
Both the setups are shown in Fig.~\ref{minitower}.
In Setup-1, four layers of tungsten and  silicon pad detector arrays are placed alternatively. 
This gives the response of incoming particles up to 4~X$_0$ in
longitudinal depth. In Setup-2, two additional tungsten layers are 
placed in front to extend the depth up to 6X$_0$.

\section{Silicon pad array and associated electronics}

\begin{figure}[tbp]
\centering
\includegraphics[width=0.7\linewidth]{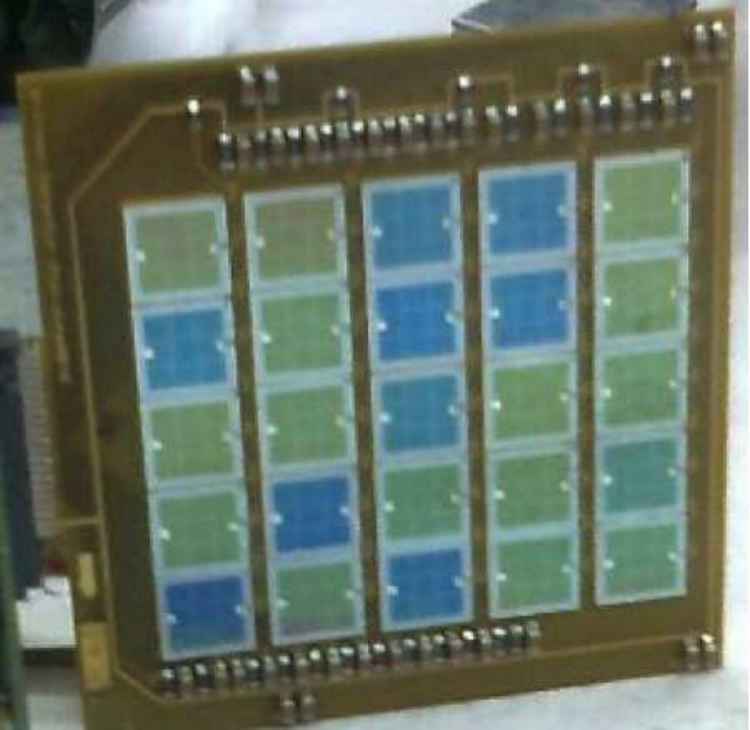}
\caption{(Color online) Sketch of the silicon detector array consisting of
$5\times 5$ single element silicon pad detectors.}
\label{silicon_pad}
\end{figure}

Each plane of the mini-tower consists of an array of $5\times 5$ single element 
silicon pads as shown in Fig.~\ref{silicon_pad}. Each silicon pad is
of 1~cm~$\times$~1~cm size and  300~$\mu$m thick.
The silicon detectors are fabricated on $\langle 111 \rangle$ 
FZ n-type wafers with $3-5$K-Ohm resistivity. The top pads are p+ and the bottom side is n+. 
The leakage current of the detectors is less than a few~nA
and breakdown voltage is above 250~V.  Each of the detectors is
surrounded by a guard-ring and is appropriately biased.   
The silicon pads have full depletion capacitance of 40~pF/cm$^2$ at 60~V.
Silicon pads are mounted on a 0.8~mm thick four layer PCB, both to hold and to connect to readout
electronics. 
The detectors were attached to the PCB with silver conductive epoxy 
of resistivity 0.006 $\Omega$-cm.
The top of the diode was coated with 1$\mu$m thick Al and 0.5$\mu$m thick
phosphosilicate glass (PSG) passivation. 
Possibility of cross talk among the silicon pad elements is negligible
because the detectors are physically  isolated from each other.

The detector signals were readout using Front End Electronics (FEE)
boards. Two different kinds of ASICs, namely, 
MANAS~\cite{muon,pmd} and ANUSANSKAR were used.
Both ANUSANSKAR and MANAS ASICs incorporate 16 pulse processing 
channels along with analog multiplexed output. Each channel is comprised
of a Charge Sensitive Amplifier (CSA), second order semi Gaussian pulse 
shapers, track and hold (T\&H) circuit 
and output buffer. For ANUSANSKAR,
AMI Semiconductor 0.7$\mu$m C07-MA technology has been used whereas
for MANAS, SCL 1.2$\mu$m C1D twin tub process have
been adopted. The size of the input transistor for both the ASICs is 
8000$\mu$m/1.5$\mu$m.
In ANUSANSKAR, CSA design is based on 
conventional folded cascode architecture with high value of feedback 
resistor, which is implemented through current conveyor method with 
improved linearity for large signals. As the input transistor plays 
a vital role in determining the noise performance of the whole amplifier, 
large area p-type MOS transistor is used as the input device to reduce the 
flicker noise. The input transistor is biased in sub threshold region with 
500~$\mu$A bias current, ensures appropriate transconductance (g$_m$) 
to reduce the contribution due to thermal noise. Semi-Gaussian 
shaping is implemented using 2$^{\rm nd}$ order Sallen-key filter 
with wide swing Operational Transconductance Amplifier (OTA)
and with a peaking time of 1.2~$\mu$s.  However, semi Gaussian shapers 
inside MANAS are implemented through Gm-C filter topology. 
The baseline recovery better than 
1\% after 4~$\mu$s in ANUSANSKAR can be achieved by tailoring the pole-zero locations 
by external DC voltage control. The T\&H block is 
used to sample the peak information of the shaped signal. 
All the 16 channel 
outputs of both the ASICs can be readout serially via analog 
multiplexer with a clock rate of 1~MHz. 
Performance summary of both the 
ASICs are given in Table~1.

\begin{table}[htbp]
\begin{center}
\footnotesize
\begin{tabular}{|c|c|c|}
\hline 
Specification & ANUSANSKAR & MANAS \\ \hline\hline
Noise at 0 pF &   390 rms electrons & 500 rms electrons    \\ \hline
Noise slope    &  7e$^-$/pf  & 11.6 e$^-$/pf    \\ \hline
Linear dynamic range & +/- 600 fC & + 500 fC to -300 fC \\ \hline
Conversion gain &  3.3 mV/fC & 3.2 mV/fC \\ \hline
Peaking time      & 1.2 $\mu$s & 1.2 $\mu$s \\ \hline
Baseline recovery & 1\% after 4 $\mu$s & 1\% after 5 $\mu$s \\ \hline
VDD/VSS              & +/- 2.5 V & +/- 2.5 V \\ \hline
Analog readout speed & 1 MHz & 1 MHz \\ \hline
Power consumption    & $\sim$15 mW/channel & ~9 mW/channel \\ \hline
Die area & 4.6 mm $\times$ 4.6 mm &  4.6 mm $\times$ 2.4 mm  \\ \hline
Technology & 0.7 $\mu$m standard  & 1.2 $\mu$m standard \\
 &  CMOS &  CMOS \\ \hline
Package   & CLCC-68 & TQFP-48 \\ \hline
\end{tabular}
\caption{Specifications of the readout ASICS (MANAS and ANUSANSKAR) 
used in the mini-tower setup and tests.}
\end{center}
\end{table}

\section{Tests of silicon pad detector with $\beta$-source}

\begin{figure}[tbp]
\centering
\includegraphics[width=0.4\textwidth]{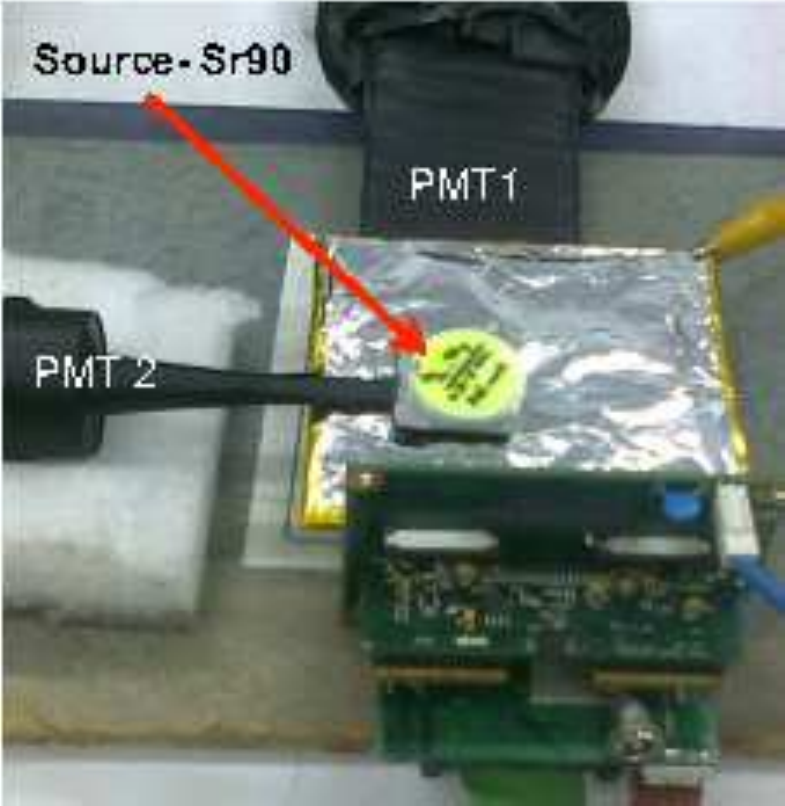}
\caption{(Color online) Laboratory setup for the test setup of silicon
  detector array with $^{90}$Sr source.}
\label{labsetup}
\end{figure}

\begin{figure}[tbp]
\centering
\includegraphics[width=0.4\textwidth]{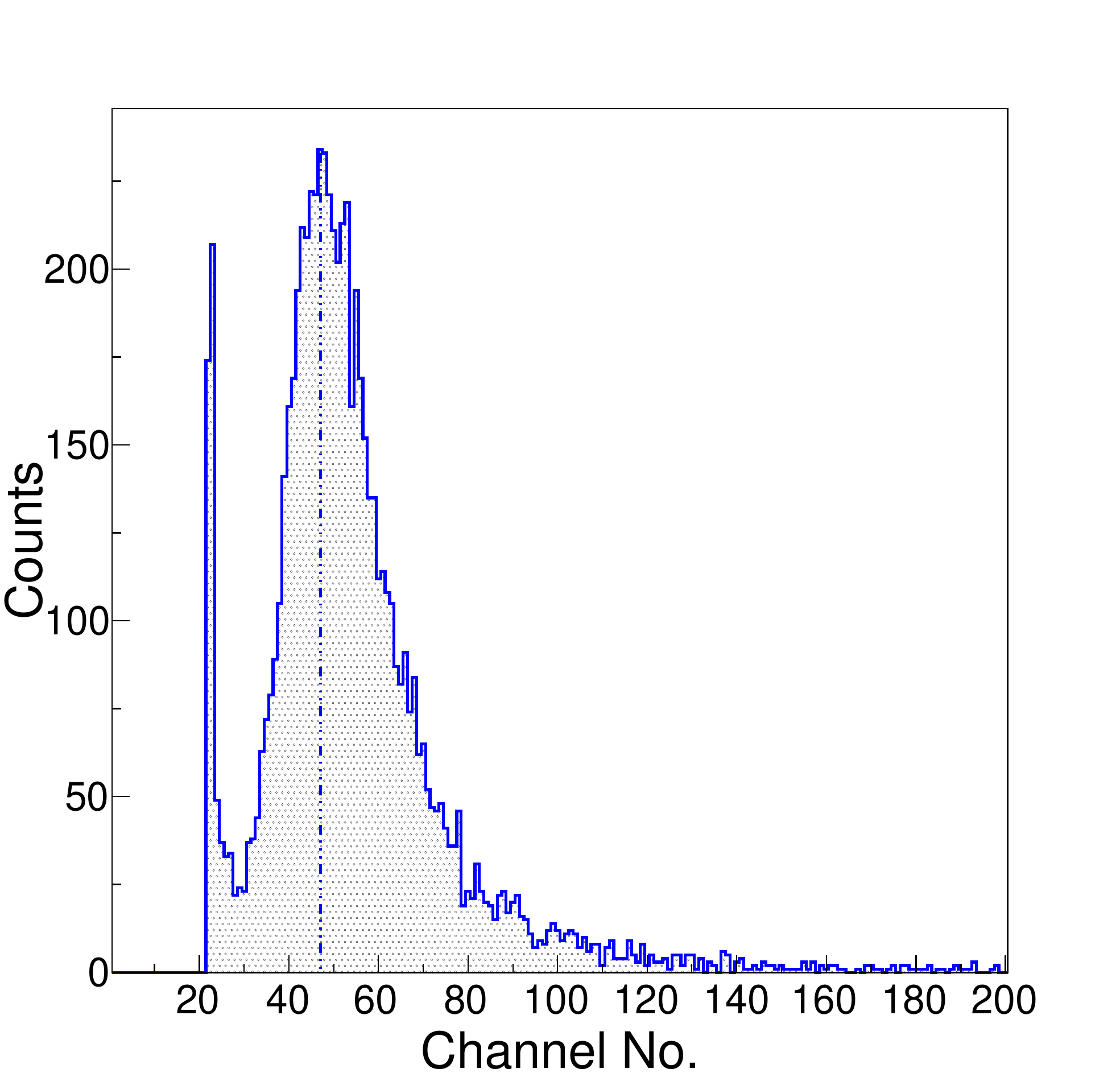}
\caption{(Color online) Response of a single silicon pad detector to $^{90}$Sr source. A
 clear peak corresponding to $\beta$~energy of 0.546~MeV is
 visible on the right, well separated from the noise (see on on the left).}
\label{beta_spectrum}
\end{figure}

Each layer of silicon detector array was tested in the laboratory 
with $^{90}$Sr $\beta$-source (with end point energy of 0.546~MeV). The test setup 
is shown in Fig.~\ref{labsetup}.  The optimum operating voltage of the detector has been 
determined after performing a voltage scan. It has been observed that the detector
can be operated at 60~Volts with reasonably good signal to noise ratio and achieved
full depletion. The source is placed on top of one of the triggering
scintillator. A
two-fold coincidence logic is derived using two scintillator paddles
as shown in the Fig.~\ref{labsetup}. 
Detector response using conventional NIM electronics for each silicon 
pad element has been carried out extensively. Figure~\ref{beta_spectrum} shows
typical response for a single silicon pad detector.
Part of the noise peak is seen on the left and the signal from the
$\beta$-source is clearly seen on the right.
This elucidates the good functionality of each element of the silicon
detector array.
To understand the exact detector response and energy loss,
tests with pion and electron beams have been performed.

\section{Test Beam Setup}

\begin{figure}[tbp]
\centering
\includegraphics[width=0.9\linewidth]{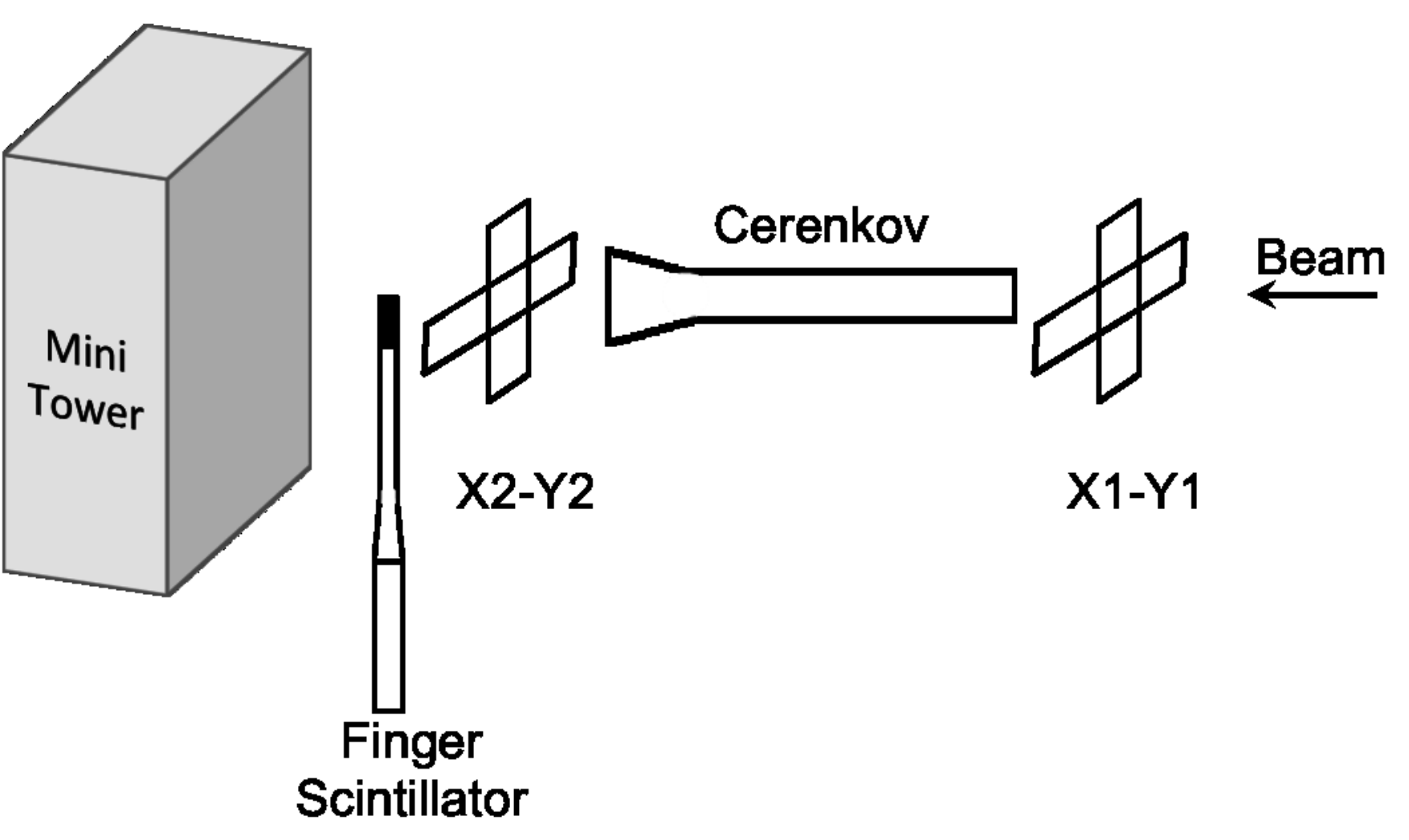}
\caption{(Color online) Sketch of the detector setup of silicon and tungsten layers.}
\label{testbeam}
\end{figure}

\begin{figure}[htbp]
\centering
\includegraphics[width=0.45\textwidth]{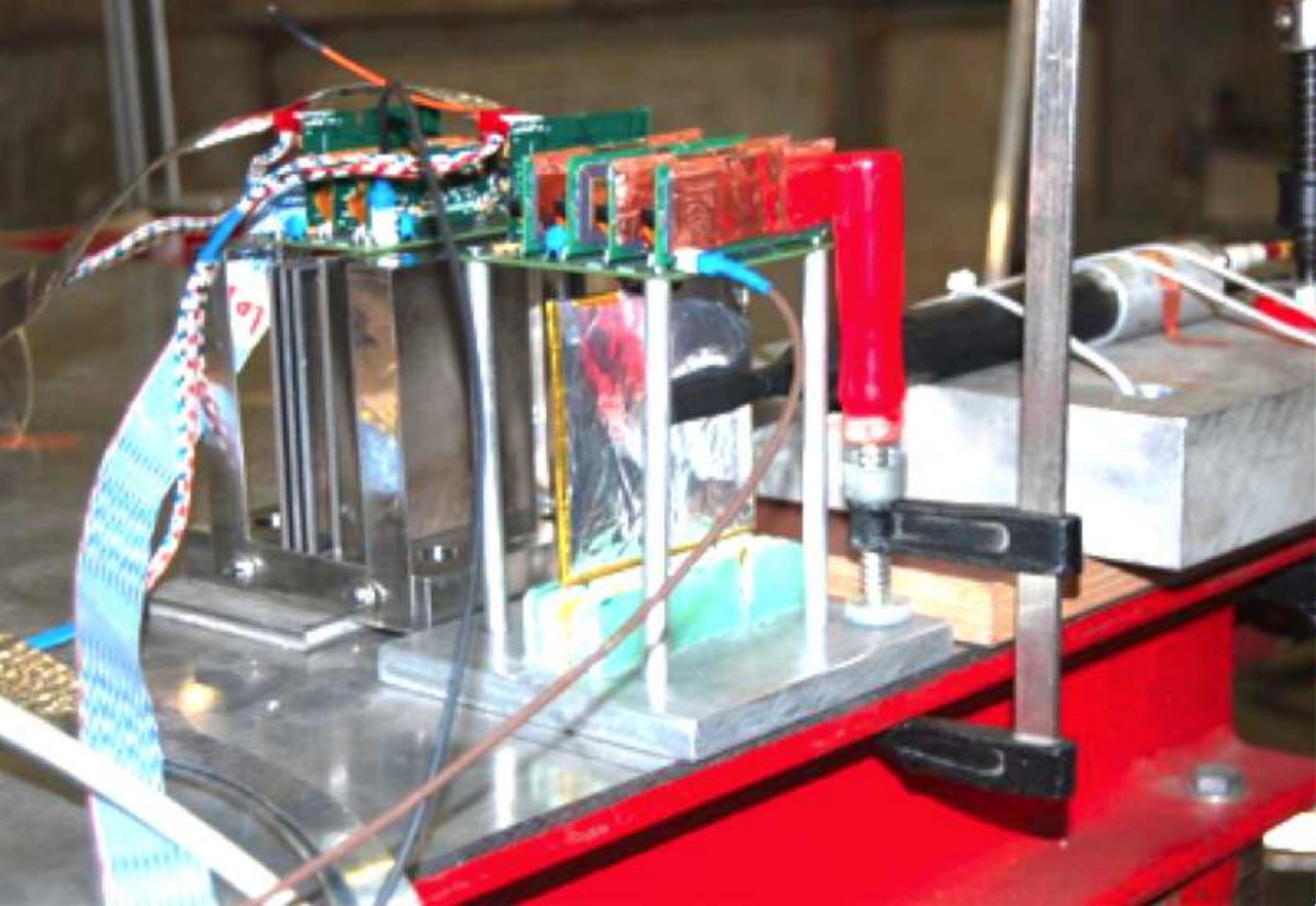}
\caption{(Color online) Photograph of the experimental Setup used in T10 PS Beam Line facility at CERN}
\label{photo}
\end{figure}

Experimental studies with the mini-tower arrangement has been carried out at the
T10 beam line of CERN-PS. The T10 beam is a secondary beam that delivers secondary particles
up to 7~GeV/c at a production angle of 61.6 milli-radians.
The secondary particles predominantly contain negative pions with a
mixture of electrons. 
A dedicated
triggering system was used to select either pion or electron beam.
The trigger system consists of two pairs of scintillator paddles,
a finger scintillator and a Cherenkov detector. The arrangement of the
trigger detectors with respect to the mini-tower is shown in
Fig~\ref{testbeam}. 
The two pairs of paddle scintillators are arranged to determine the 
$x-y$ positions of the incoming beam within 1~cm$^2$ area. 
The pion trigger is generated using the scintillator paddles in
coincidence with a 
small finger
scintillator of size 3~mm~$\times$~3~mm, placed close to the mini-tower.
Additionally, a
Cherenkov detector was used for electron trigger. The gas pressure of
the Cherenkov detector was adjusted to obtain highest purity of the
electrons. 

The mini-tower assembly was placed on a movable table to adjust
detector position with respect to the beam. 
A photograph of the detector setup in the T10 beam line is shown in Fig.~\ref{photo}.
The silicon pad arrays, along with backplane PCBs are properly shielded against EMI and ambient
light for better signal to noise ratio.
The detector signals are readout by using a FEE  board and 
further processed by a MARC~ASIC, which communicates with the
Cluster Read Out Concentrator Unit System (CROCUS)~\cite{muon,pmd}.
Finally, the CROCUS interfaces to the data acquisition system via
fibre optic cable.

\section{Results and discussion}

\begin{figure}[tbp]
\begin{center}
\includegraphics[scale=.34]{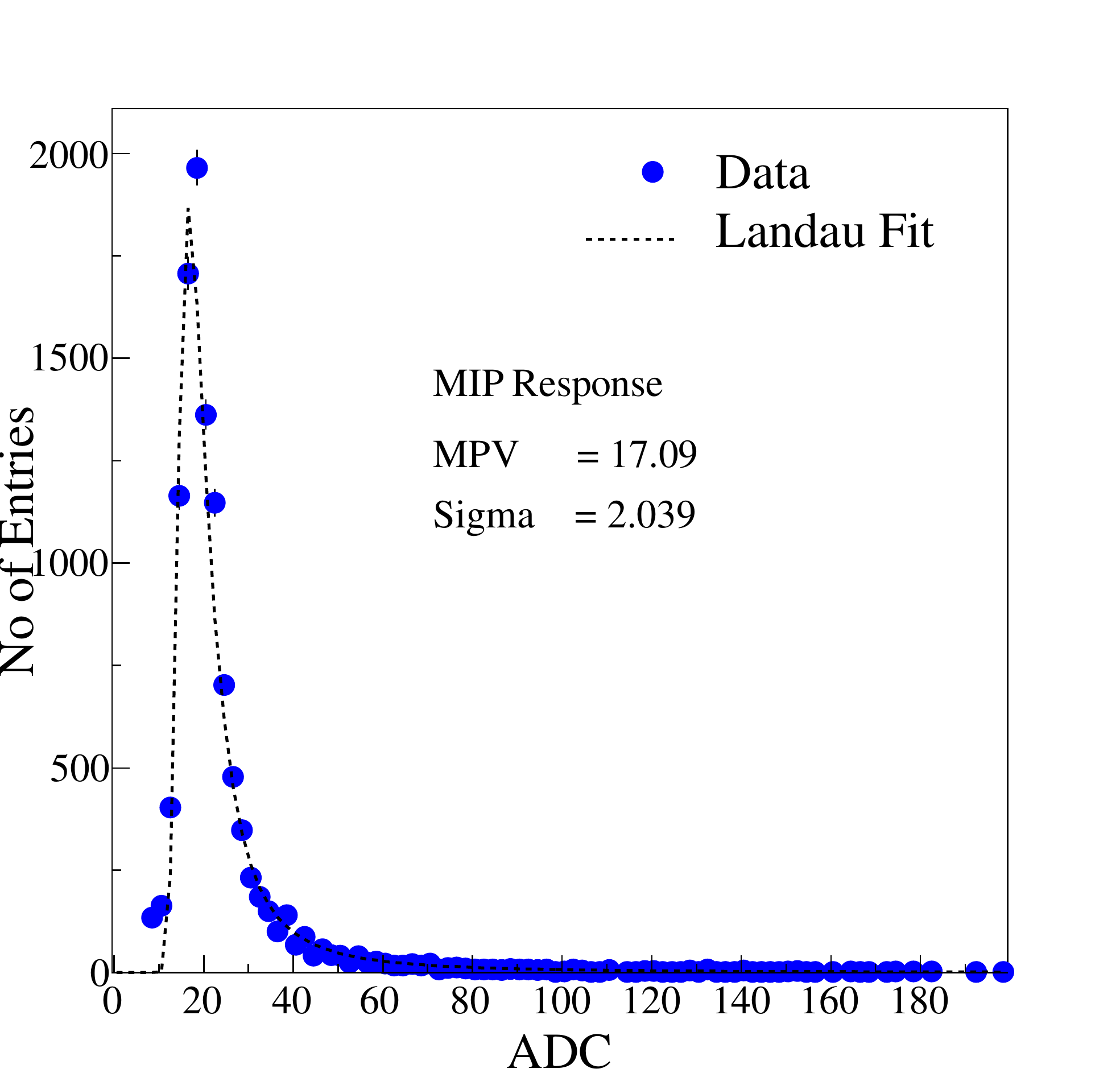}
\includegraphics[scale=.34]{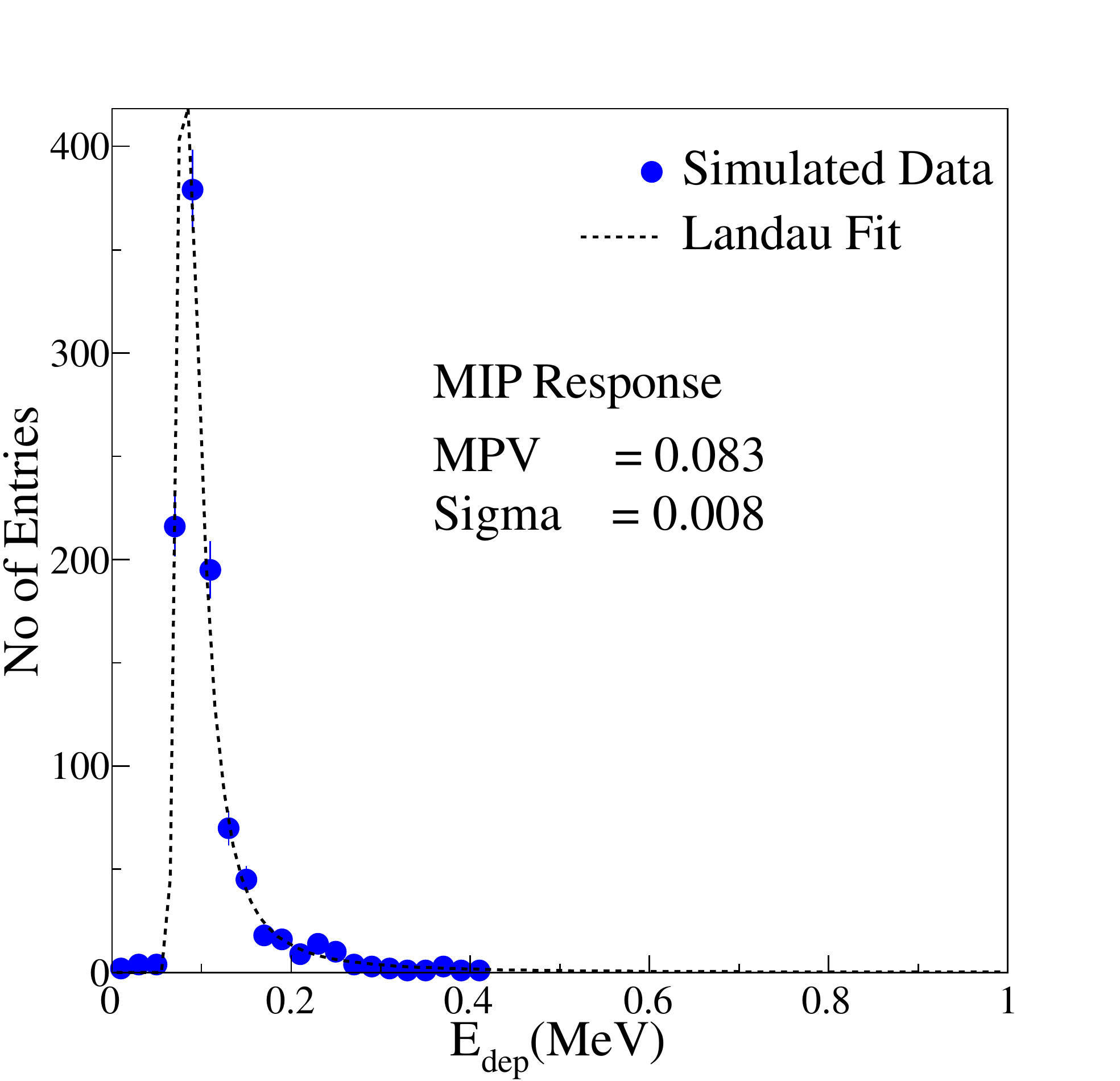}
\caption{(Color online).
Response of silicon pad detector arrays to pions, showing the distribution 
similar to the minimum ionizing particles: in ADC for  test beam data
(upper panel) and in MeV from simulated data (lower panel).
}
\label{pion}
\end{center}
\end{figure}

\begin{figure}[tbp]
\begin{center}
\includegraphics[scale=.5]{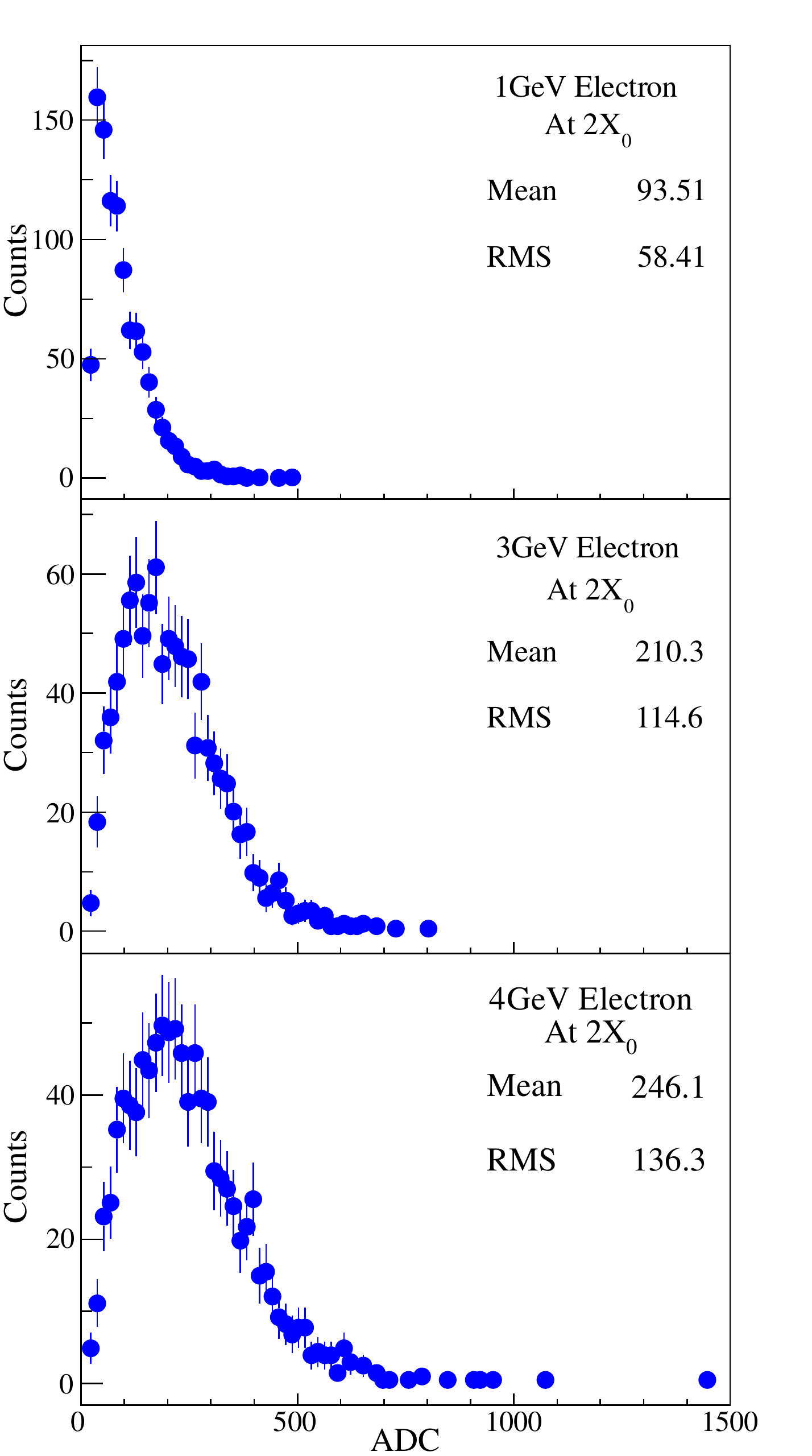}
\caption{(Color online).
Response of silicon pad detector arrays for 
electrons at three different energies, after 2$X_0$ thick tungsten absorber.
}
\label{electron}
\end{center}
\end{figure}

The test beam results for incident pions and electrons with different
mini-tower setups are presented here and compared to those from the
simulated data.
The simulation takes care of each and every aspect of the actual
experimental setup.
The gap between two tungsten layers is kept as 0.21~cm,
out of which 0.03~cm is for the silicon wafer followed by 0.08~cm
thick PCB and the rest 0.1~cm is kept for associated electronics. The
PCB and electronics parts are included in the simulation.
The simulated data gives the 
energy deposition in each silicon pad for different incoming
beams. Responses from each layer of the detector 
has been extracted by summing up the signals from each layer.
Performances of the detector using both the setups of the mini-tower 
to photons, electrons 
and neutral pions of different energies were studied extensively using the
simulated data.

\begin{figure}[tbp]
\centering
\includegraphics[scale=0.34]{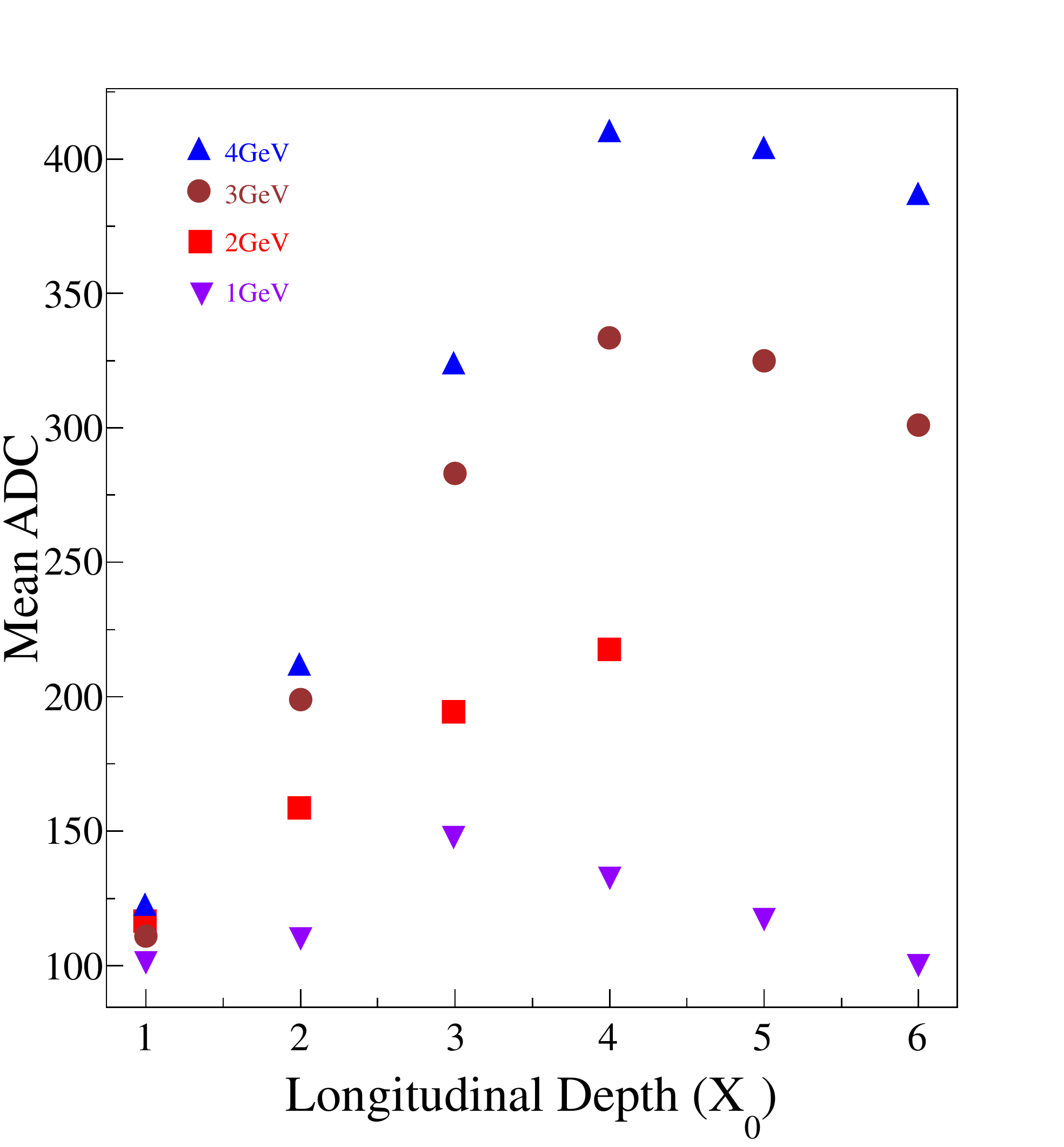}
\includegraphics[scale=0.34]{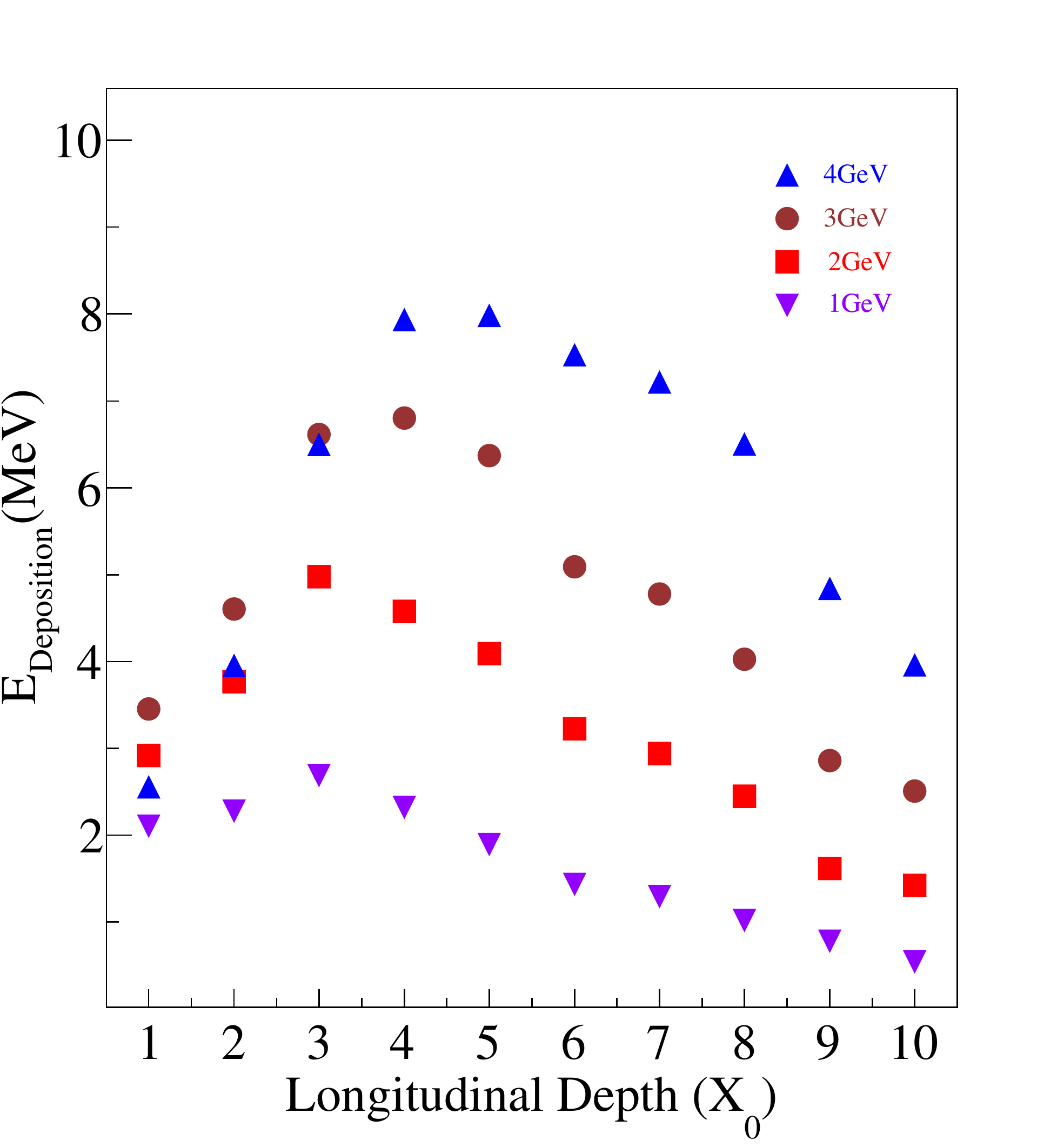}
\caption{(Color online) Longitudinal shower profile using experimental
data (upper panel) and simulated data (lower panel).}
\label{long_data}
\end{figure}

\begin{figure}[tbp]
\begin{center}
\includegraphics[scale=0.34]{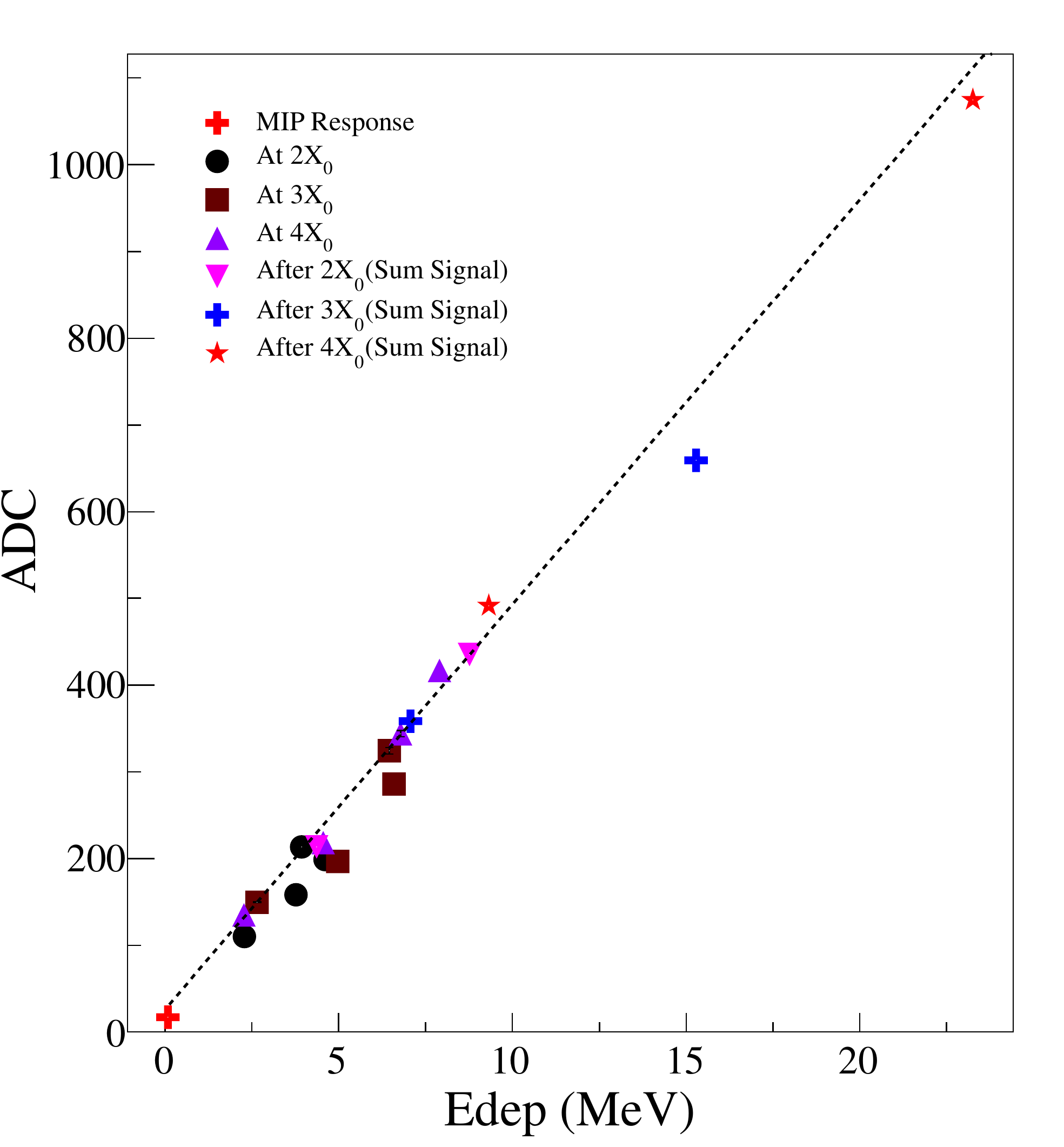}
\caption{(Color online) Conversion Curve from experimentdal data (ADC)
to energy deposition from simulated data (MeV).}
\label{conversion}
\end{center}
\end{figure}

The mini-tower was exposed to pion and electron beams. 
Due to lighter mass and electromagnetic
nature, electrons produce shower~\cite{EM-shower-1} which propagate through
the layers of the tower and leave its footprints for offline
reconstruction. Pions are
less likely to produce shower within the depth of the calorimeter,
and behave like minimum ionizing
particles (MIP).  The energy deposition of the 
MIPs can be explained by the Landau
Distribution.
Analysis of test beam data shows that pions indeed behave like MIP
with a most probable value (MPV) of 17~ADC, as shown in the upper
panel of Fig.~\ref{pion}. 
This corresponds to conversion gain of 3.24~mV/fC of the ASIC used.
The results of the simulation for pions, shown in the lower panel of
Fig.~\ref{pion} show a similar nature of the MIP spectrum with
a MPV value of 83~keV.

Electrons produce electromagnetic shower in the mini-tower.
The shower shape and energy deposition in each silicon layer gives
an important input for optimizing the granularity
along with transverse and longitudinal leakage of energy in the calorimeter. 
Figure~\ref{electron} gives the energy deposition of electromagnetic
showers produced after 2$X_0$ for three different electron energies.
It is observed that the shower profile is not so well developed for 1~GeV
electron, but as the electron energy increases the distribution becomes more
towards Gaussian. The mean of the distributions shift with the
increase of energy, as expected.

For different incident energy of electrons, energy depositions are
calculated for different silicon pad detector arrays
as a function of longitudinal depth from 1$X_0$ to 6$X_0$.
Figure~\ref{long_data} shows the mean energy deposition 
as a function of longitudinal depth
of the mini-tower. The experimental data, presented in the upper panel, 
shows that with the increase of the longitudinal depth, the energy
deposition for a given incident energy increases at first and 
then comes to a maximum. Simulated results, as shown in the lower
panel of the figure, confirms to the experimental findings. These show that the energy
deposition from the resulting shower increases with the increase of
depth of the calorimeter until it attains a maximum (shower max) and
then deteriorates. The position of the 
shower max is found to shift to larger depth with
increase in the energy of the incident particle. 
The shower profiles observed in
the experimental data are similar to what is expected from the simulation.
This is reassuring in the process of making a full calorimeter.

The energy depositions for all combinations of absorber material
thickness and beam energies from simulations and the experimental data
have been calculated and plotted in Fig.~\ref{conversion}. The
abscissa gives the energy deposition from simulated data whereas the
ordinate shows the experimental observations. 
For the electron
data, signals corresponding to each layer as well as sum of signals from
different layers are presented.
MIP result is also superimposed on the same figure. 
A linear relationship is 
observed over a wide range of energy depositions. A linear fit through
the data points yields, ADC = 46.7~$\times$E$_{dep}$ + 26, where
E$_{dep}$ is expressed in MeV.
The offset maybe because of the electronics conversion and gain.
This assures 
the quality of the measurement and will be useful while designing the
full silicon-tungsten calorimeter.

\section{Summary}

In summary, 
we report a novel design of a sampling calorimeter using
silicon pad detector arrays and tungsten converters. The performance
of the calorimeter has been studied by using detailed simulation as
well as by constructing a prototype mini-tower.
The mini-tower consisted of
four layers of silicon pad detector arrays and several tungsten
plates.  Two different configurations of the mini-towers have been used to
 probe from 1-6$X_0$ depth in the longitudinal direction.
Laboratory tests using $^{90}$Sr source were made to ensure the
good quality of the silicon pad detectors. The mini-tower was tested
using pion and electron beams from 1$-$6~GeV. Pions give response
corresponding to minimum ionizing particles, whereas electrons produce
shower. The energy deposition resulting from the showers have been
measured and compared with results from simulations. A linear relationship between
the energy depositions in the simulation and data was obtained, which
ensures satisfactory performance of the silicon-tungsten mini-tower.

\vspace{0.5cm}

\noindent
{\bf Acknowledgement}

We acknowledge the support of C.K. Pithawa, Dinesh
Srivastava, Yogendra Viyogi, R.K. Bhandari, Werner Riegler and Paolo
Giubellino.
We thank the ALICE FoCaL collaboration, especially
discussions with Premomoy Ghosh, Taku Gunji, Marco Van Leeuwen, Gert-Jan Nooren, Elena Rocco
and Thomas Peitzmann.
We acknowledge Bharat Electronics Limited, Bangalore for
providing good quality silicon detectors.
We thank the
CERN PS team for providing excellent quality beam for the detector tests.

\end{document}